\documentclass[fleqn,9pt,onecolumn]{wlscirep}
\usepackage[utf8]{inputenc}
\usepackage{hyperref}
\usepackage[T1]{fontenc}
\usepackage{helvet}

\title{Particle-hole asymmetry in the  dynamical spin and charge structure factors of the corner-shared one-dimensional cuprates}

\author[1,*]{Shaozhi Li}
\author[2,3]{Alberto Nocera}
\author[4]{Umesh Kumar}
\author[5,6,$\dagger$]{Steven Johnston}
\affil[1]{Materials Science and Technology Division, Oak Ridge National Laboratory, Oak Ridge, Tennessee 37831, USA}
\affil[2]{Stewart Blusson Quantum Matter Institute, University of British Columbia, Vancouver, British Columbia, Canada V6T 1Z4}
\affil[3]{Department of Physics Astronomy, University of British Columbia, Vancouver, British Columbia, Canada V6T 1Z1}
\affil[4]{Theoretical Division, T-4, Los Alamos National Laboratory, Los Alamos, New Mexico 87545, USA}
\affil[5]{Department of Physics and Astronomy, The University of Tennessee, Knoxville, Tennessee 37966, USA}
\affil[6]{Joint Institute of Advanced Materials at The University of Tennessee, Knoxville, Tennessee 37996, USA}

\affil[*]{\href{mailto:lis1@ornl.gov}{lis1@ornl.gov}}
\affil[$\dagger$]{\href{mailto:sjohn145@utk.edu}{sjohn145@utk.edu}}

\begin{abstract}
The collective spin and charge excitations of doped cuprates and their relationship to superconductivity are not yet fully understood, particularly in the case of the charge excitations. Here, we study the doping-dependent dynamical spin and charge structure factors of single and multi-orbital models for the one-dimensional corner shared spin-chain cuprates using several  numerically exact methods. We find that the singleband Hubbard model can describe the spin and charge excitations of the $pd$-model in the low-energy region, including the particle-hole asymmetry in the spin response. However, our results also reveal that the weight of the interorbital spin excitations between Cu and O orbitals is comparable to the weight of the spin excitations between two Cu orbitals. This finding elucidates the microscopic nature of the spin excitations in the 1D cuprates and sheds light on the spin properties of other oxides. Importantly, we find a particle-hole asymmetry in the orbital-resolved charge excitations, which cannot be described by the singleband Hubbard model and is relevant to resonant inelastic x-ray scattering experiments. Our results imply that the explicit inclusion of the oxygen degrees of freedom may be required to understand experimental observations.
\end{abstract}
\begin{document}

\flushbottom
\maketitle

\thispagestyle{empty}

\section*{Introduction}
The high-temperature (high-$T_c$) superconducting cuprates are governed by competition between multiple intertwined orders including unconventional superconductivity, pseudogap behaviour, and various spin- and charge-orders \cite{KeimerReview}. Determining how well these competing orders and their associated dynamical fluctuations are described by the singleband Hubbard model has become a frontier problem in condensed matter physics~\cite{LeBlancPRX2015, Jiang1424, QinPRX2020}. Nevertheless, significant progress has been made towards understanding the physics of the singleband Hubbard model, including its superconducting~\cite{QinPRX2020, MaierPRL2005, GullPRL2013, Jiang1424, JiangPRR} and normal-state transport properties \cite{MousatovPRL2019, Huang987, FedorPRL2020}, its pseudogap~\cite{Chen2017, Robinson_2019, WeiPRX2018} and stripe orders~\cite{Zheng1155, EhlersPRB2017, DarmawanPRB2018, Huangnpj2018}, and its dynamical response functions~\cite{JiaNacom2013, Ishii2014, Chen2017, LeBlancPRB2019}. While it is now clear that the predictions of the singleband Hubbard model are consistent with many experimental observations in the cuprates, recent x-ray scattering and nuclear magnetic resonance experiments have inferred that the oxygen $2p$ orbitals provide important contributions to the spin- and charge- orders~\cite{RybickiNatureComm, AchkarNatureMat, JurkutatPRB2014}. These observations raise questions on the validity of the singleband Hubbard model for describing some properties of the cuprates, including the fluctuations of their intertwined orders.

Studying the ground and excited state properties of the single- and multi-band Hubbard model on large two-dimensional (2D) lattices is challenging. For example, theoretical studies using the dynamic mean-field theory and its cluster and diagrammatic extensions~\cite{ RubtsovPRB2009, SunPRB2002, ToschiPRB2007, KataninPRB2009} have been limited to relatively small clusters. The study of the Hubbard model can be extended to larger lattices with determinant quantum Monte Carlo (DQMC)~\cite{WhitePRB1989,JiaNacom2013,Huang1161,Huangnpj2018,Huang987} but this method is more limited by the sign problem in comparison to embedded cluster methods. Some zero temperature techniques, such as the tensor network~\cite{ORUS2014117, OrusNatur2019} and the path-constrained auxiliary-field quantum Monte Carlo~\cite{QinPRB2016} can access large lattices, but have difficulty in calculating the single- and two-particle dynamical response functions probed by angle-resolve photoemission (ARPES), inelastic neutron scattering (INS), and resonant inelastic x-ray scattering (RIXS) experiments. Importantly, contrasting results obtained from the singleband model with multi-orbital models adds another layer of difficulty, since the inclusion of the O $2p$ orbitals significantly increases the complexity of the problem~\cite{FrickPRB1990, Avella2013, Huang1161, MaiQuantMat2021}.

In this work, we study single- and multi-orbital models for quasi-one-dimensional (1D) corner-shared spin-chain cuprates like Sr$_2$CuO$_3$ and the recently synthesized doped chains Ba$_{2-x}$Sr$_x$CuO$_{3+\delta}$ \cite{PhysRevMaterials.3.044802}. 
Specifically, we compute and contrast the momentum-resolved dynamical spin and charge responses in these models using DQMC, density matrix renormalization group (DMRG), and exact diagonalization (ED, see Supplementary Note 1). By focusing on 1D cuprate models, we are able to perform reliable calculations for the single- and two-particle response functions for large system sizes with good momentum resolution, even in the multi-orbital case. These calculations are enabled by the fact that 1D systems generally have manageable fermion sign problems\cite{ShaozhiPRB2018} and algorithmic advances in computing dynamical response functions using DMRG \cite{PhysRevB.60.335, PhysRevE.94.053308}. 

\section*{Results}

\subsection*{Corner-Shared Spin Chains} 
The active orbitals in spin-chain cuprates like 
$\mathrm{Sr}_2\mathrm{CuO}_3$ and Ba$_{2-x}$Sr$_x$CuO$_{3+\delta}$ are located in their $\mathrm{CuO}_4$ plaquettes, which are arranged in a corner-shared geometry, as shown in Fig.~\ref{Fig:fig1}{\bf a}. $\mathrm{Sr}_2\mathrm{CuO}_3$ has been extensively studied both experimentally~\cite{AmiPRB1995, KojimaPRL1997, Walters2009, PhysRevLett.93.087202, SchlappaNature2012, Schlappa2018} and using singleband models like the Heisenberg model, the $t-J$ model~\cite{AmiPRB1995, KojimaPRL1997, Kumar_2018, Schlappa2018, Walters2009, PhysRevLett.93.087202}, and the extended Hubbard model~\cite{NeudertPRL1998, PencPRB2000}. These models generally provide an excellent description of the magnetic excitations -- in this case a multi-spinon continuum --  observed in INS~\cite{AmiPRB1995, KojimaPRL1997, Walters2009, PhysRevLett.93.087202} and RIXS \cite{SchlappaNature2012, Schlappa2018} experiments. The spin and charge excitations of doped samples have received comparatively less attention because doping electrons or holes in $\mathrm{Sr}_2\mathrm{CuO}_3$ has proved to be challenging~\cite{MaitiPRB2002, PhysRevLett.111.067204,  PhysRevB.95.235154}. However, the Ba$_{2-x}$Sr$_x$CuO$_{3+\delta}$ system offers new possibilities in this regard\cite{PhysRevMaterials.3.044802}. 

To determine the influence of the orbital degrees of freedom on the spin and charge dynamics of the cuprates, we consider a four-orbital $pd$-model for corner-shared cuprates, which includes the Cu $3d_{x^2-y^2}$ and O $2p_{x/y}$ orbitals near the Fermi level, (Fig.~\ref{Fig:fig1}{\bf a}). To isolate the influence of the oxygen degrees of freedom, and assess the validity of a singleband effective model, we also studied a singleband $t$-$t^\prime$ Hubbard model with model parameters selected to reproduce results from our multi-orbital model (see Methods and Supplementary Note 2). 

\subsection*{Electronic Structure} 
\begin{figure*}[ht]
\centering
\includegraphics[width=0.8\columnwidth]{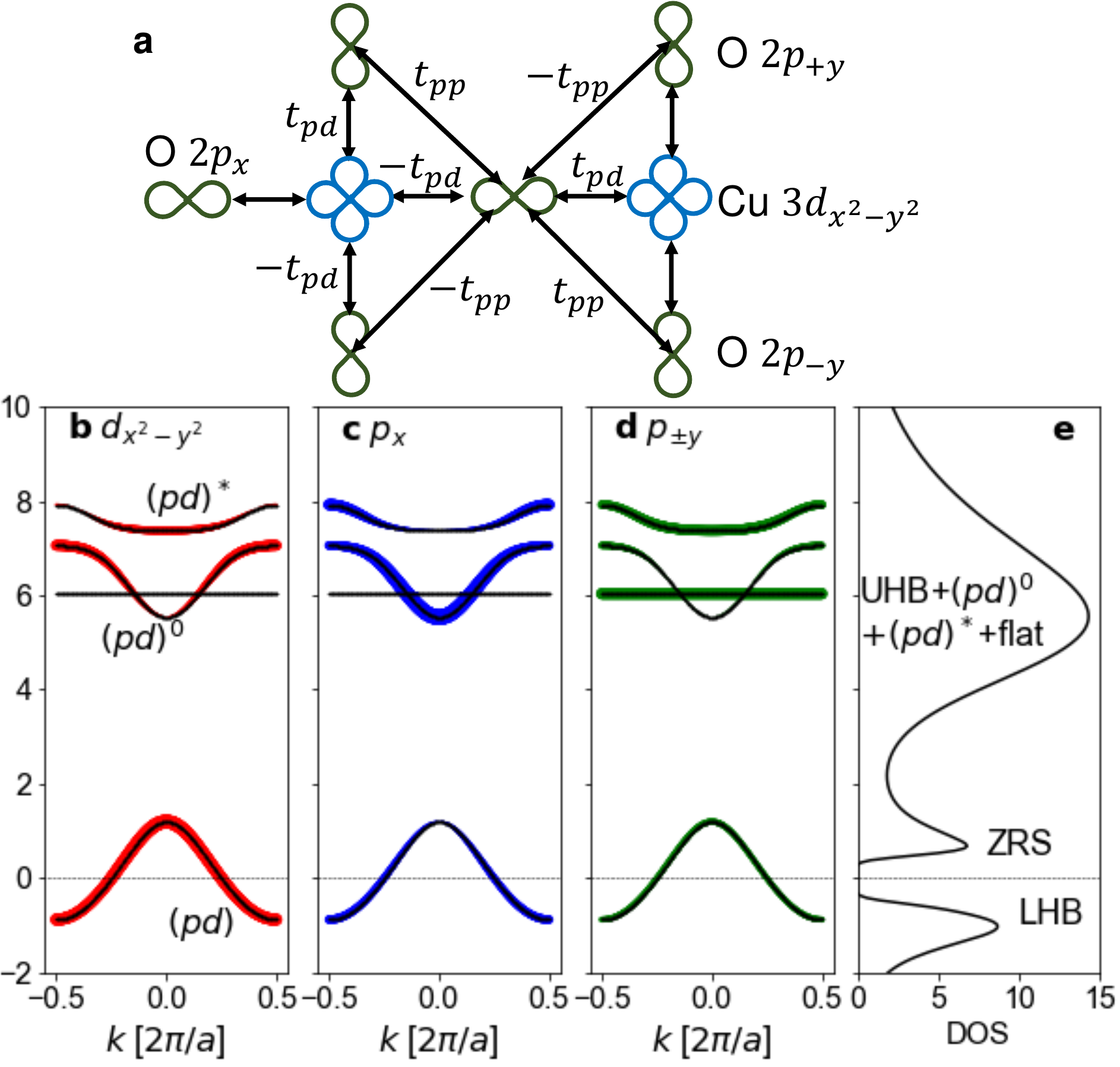}
\caption{\textbf{The multi-orbital model for the corner-shared cuprate spin chains.} {\bf a} A sketch of the four-orbital $pd$-model describing the corner-shared spin-chain cuprates like Sr$_2$CuO$_3$. Panels {\bf b}-{\bf d} plot the noninteracting band structure in hole language and at half-filling. In each panel, the weight of the Cu $3d$ and O $2p_x$ and $2p_{\pm y}$ orbitals are indicated by the weight of the colored overlays. Panel {\bf e} shows the density of states (DOS) of the $pd$-model with interactions at half-filling, obtained from DQMC calculations at $T = 0.0625$ eV. Here, LHB (UHB) and ZRS denote the portions of the spectra corresponding to the lower (upper) Hubbard band and the Zhang-Rice Singlet quasi-particle band, while ``flat" indicates portions of the electronic structure arising from the non-bonding oxygen $2p_y$ flat band appearing in panel {\bf d}.}
\label{Fig:fig1}
\end{figure*}

There is a significant orbital overlap between the Cu and O orbitals in quasi-1D and 2D cuprate materials, which hybridizes the Cu $3d_{x^2-y^2}$ and O $2p_{x,y}$ orbitals. In the non-interacting limit, these orbitals form bonding $(pd)$, nonbonding $(pd)^0$, and anti-bonding $(pd)^*$ bands, as shown in Figs.~\ref{Fig:fig1}{\bf b} - \ref{Fig:fig1}{\bf d}. In the corner-shared cuprates, there is also a flat band, which originates from a nonbonding combination of $2p_{\pm y}$ orbitals that do not hybridize with the $3d_{x^2-y^2}$ orbital. 

\begin{figure*}[ht]
    \centering
    \includegraphics[width=0.5\columnwidth]{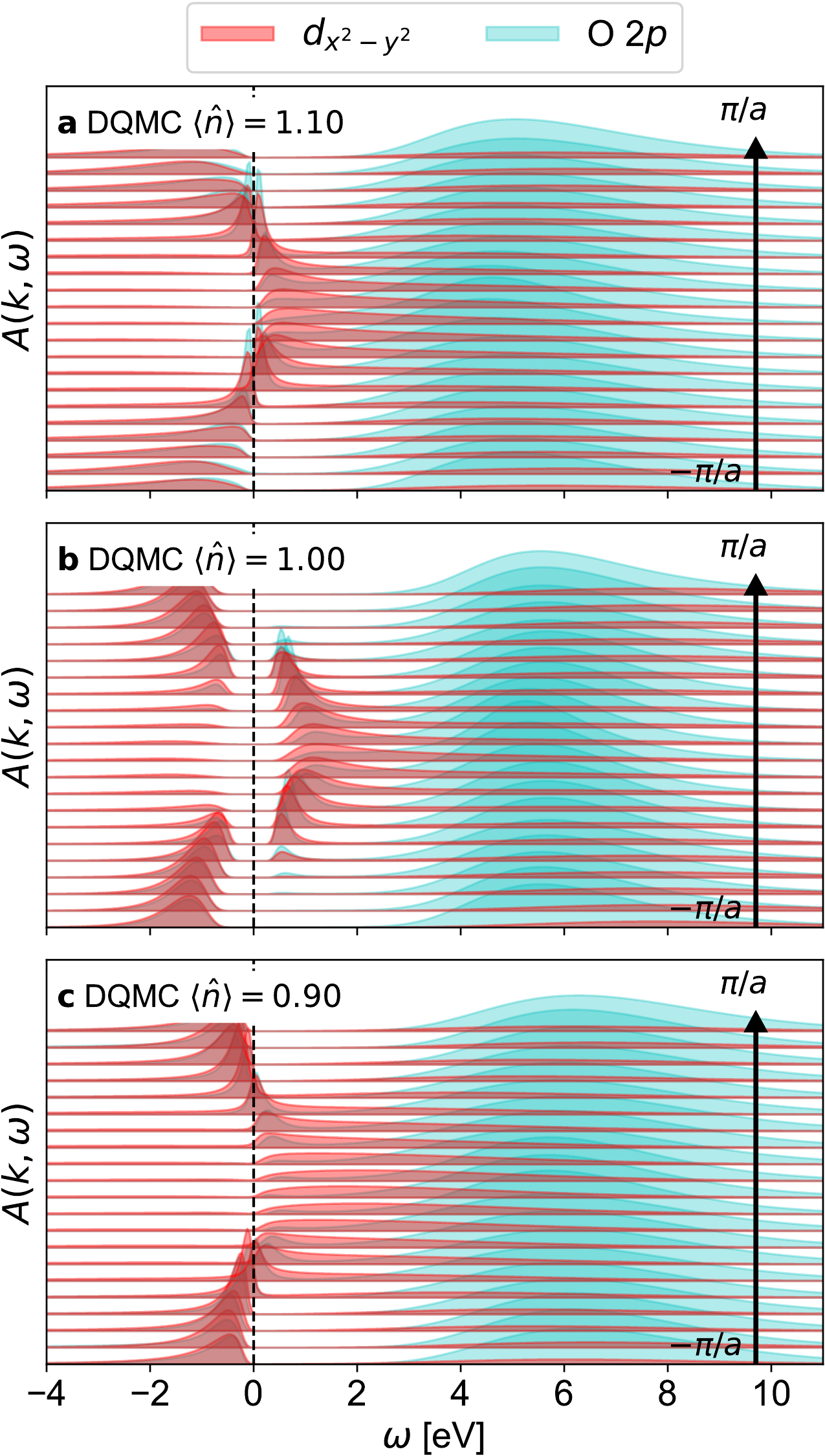}
    \caption{\textbf{The single electron spectral function $A(k,\omega)$ of the multi-orbital $\mathbf{pd}$-model computed using DQMC.} Results are shown for $\langle n \rangle = 1.1$, $1$, and $0.9$ holes/Cu.  Red color represents Cu $d_{x^2-y^2}$, while cyan color represent the sum of O $p_x$ and O $p_{\pm y}$ components. From bottom to top, the momentum $k$ increases from $-\pi/a$ to $\pi/a$ in each panel. In both cases, results were obtained on a chain with $N=20$ unit cells. 
     }
    \label{fig:Akw_DQMC}
\end{figure*}

\begin{figure*}[ht]
    \centering
    \includegraphics[width=0.5\columnwidth]{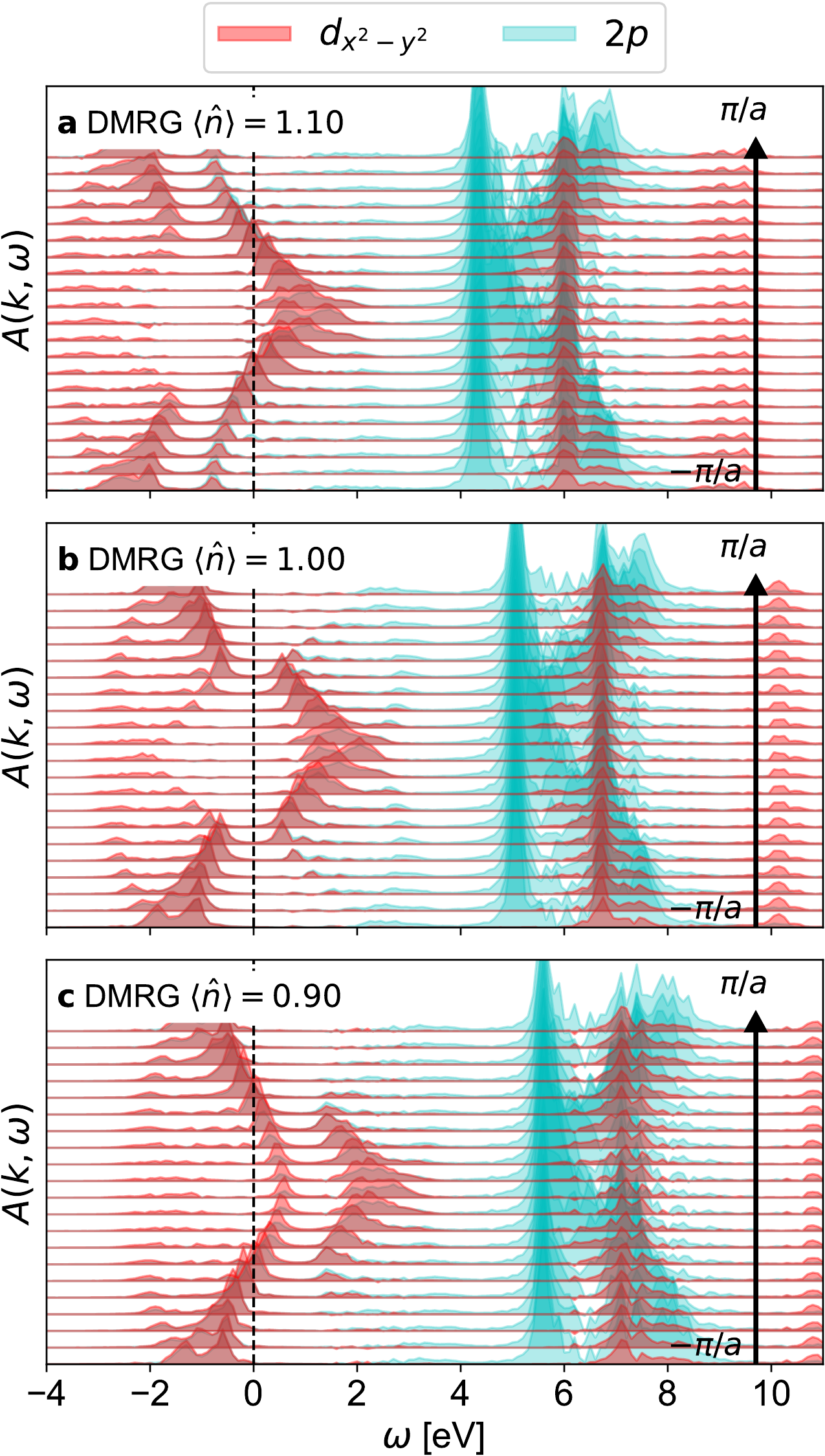}
    \caption{\textbf{The single electron spectral function $A(k,\omega)$ of the multi-orbital $\mathbf{pd}$-model computed using DMRG.} Results are shown for $\langle n \rangle = 1.1$, $1$, and $0.9$ holes/Cu.  Red color represents Cu $d_{x^2-y^2}$, while cyan color represents the sum of O $p_x$ and O $p_{\pm y}$ components. From bottom to top, the momentum $k$ increases from $-\pi/a$ to $\pi/a$ in each panel. In both cases, results were obtained on a chain with $N=20$ unit cells. 
     }
    \label{fig:Akw_DMRG}
\end{figure*}

Throughout, we work in the hole-language where $\langle \hat{n} \rangle~=~1$ corresponds to half-filling (i.e. 1 hole/Cu), and $\langle\hat{n} \rangle > 1$ ($< 1$) corresponds to hole (electron) doping. At half-filling, when the interactions are turned on, the bonding $(pd)$ band is split into the lower and upper Hubbard band (LHB/UHB) as well as an additional Zhang-Rice band (ZRS) located above the Fermi level~\cite{ZhangPRB1988,TjengPRL1997}. These bands can be easily resolved in both the single-particle spectral functions shown in Figs.~\ref{fig:Akw_DQMC}{\bf b} and  \ref{fig:Akw_DMRG}{\bf b}, and in the total interacting density of states (DOS) shown in Fig.~\ref{Fig:fig1}{\bf e}. For our parameters, the energy of the upper Hubbard band is close to the O-derived bands, leading to a broad peak in the DOS centered around $\omega=6$ eV. 

To study the electronic structure of the interacting $pd$-model, we plot the momentum-dependent spectral functions $A(k,\omega)$ in Figs.~\ref{fig:Akw_DQMC} and \ref{fig:Akw_DMRG} for hole concentrations $\langle \hat{n}\rangle=1.1$, 1, and 0.9. Figures~\ref{fig:Akw_DQMC} and  ~\ref{fig:Akw_DMRG} show results obtained from finite-temperature DQMC and zero temperature DMRG calculations, respectively. From bottom to top, the momentum $k$ increases from $-\pi/a$ to $\pi/a$ in each panel. The black dashed line represent the Fermi level $E_\mathrm{F}$. To highlight the orbital content of the spectral features, we indicate the Cu $d_{x^2-y^2}$ orbital of the spectral function with red color and the sum of the spectral weights of the O $p_x$ and O $p_{\pm y}$ components in cyan. The zero-temperature spectral function shown in Fig.~\ref{fig:Akw_DMRG} exhibits sharper features and hence richer details compare to the DQMC results.

For simplicity, we begin by discussing the spectra at half-filling. Both DQMC and DMRG have a clear gap at the Fermi level $E_\mathrm{F}$, consistent with a Mott-insulating state. In the low energy region $[-4,4]$~eV, the DMRG spectra also have footprints of spin-charge separation, consistent with ARPES measurements on SrCuO$_2$~\cite{Kim2006}.
Besides, the DMRG results show a dispersing band between 6 and 8~eV with a significant amount of O $2p$ character and two flat bands at $\omega=6$~eV and $\omega=7.2$~eV. The flat band at lower energy is composed almost entirely of the $O$ 2$p_{\pm y}$ orbitals and corresponds to the flat band shown in the noninteracting band structure. The higher-energy flat band is mixed between Cu and O and corresponds to the UHB with additional weak Cu satellites at $\omega\approx 10$~eV. Note that the contribution of the Cu orbitals to the anti-bonding $(pd)^{*}$ band is reduced significantly in the interacting case compared to the noninteracting case. Due to a combination of thermal broadening and the use of the Maximum Entropy method, these fine structures blend into a broad peak in the DQMC results in the same energy region.

By considering the full $pd$-model, we can access the Cu and O components to the spinon and holon states in the low energy region, which have not been reported in the literature to our knowledge. Our DMRG results show that the Cu and O weights of the main structures are comparable over the entire Brillouin zone. We note, however, that the holon-shadow bands are dominated by the Cu component below the Fermi level near $k=0$ but dominated by the O component above the Fermi level near $k=\pm \pi/a$. Our DQMC results show a similar composition for the main structures, but the intensity of the shadow bands is too weak to be captured by our Maximum Entropy method.

The system undergoes a metal-insulator transition as the system is either electron- or hole-doped. In our $pd$-model, we see that spinon-antiholon branches are responsible for the spectral weight crossing the Fermi level, consistent with the results of the singleband Hubbard model~\cite{PhysRevB.97.045146}. The orbital components in both low and high energy regions do not change much compared to the half filling case.

\subsection*{The Magnetic Excitations} 

\begin{figure*}[ht]
\centering
\includegraphics[width=\textwidth]{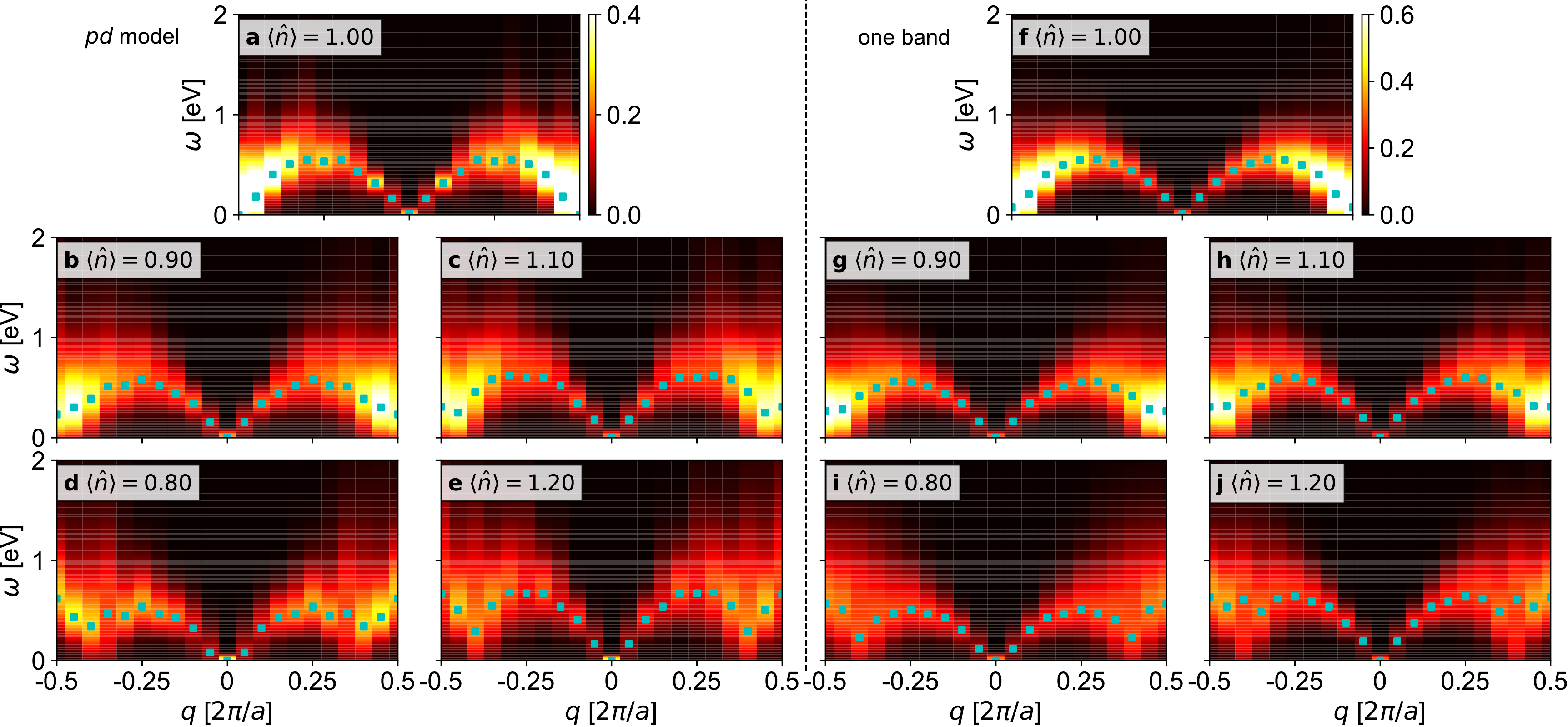}
\caption{\textbf{Finite temperature DQMC results for the total dynamical spin structure factor of the multi-orbital $\mathbf{pd}$- and singleband Hubbard models.} Panels {\bf a}-{\bf e} show the dynamical spin structure factor $S(q,\omega)$, obtained from DQMC simulations of the full multi-orbital $pd$-model at various fillings, as indicated. Panels {\bf f}-{\bf j} show corresponding results for DQMC simulations of the singleband Hubbard model. Both sets of results were obtained using chains with $N = 20$ unit cells and at temperature $T = 0.0625$ eV.}
\label{Fig:DQMC_spin}
\end{figure*}

We now examine the collective magnetic excitations of the $pd$-model. Figure \ref{Fig:DQMC_spin} summarizes our finite-temperature ($T = 0.0625$ eV) DQMC results for the dynamical spin structure factor $S(q,\omega)$ for doping levels spanning from $\langle \hat{n} \rangle = 0.8 - 1.2$. (Additional DQMC data for the  doping levels not shown here are provided in Supplementary Note 3.)   Figure~\ref{Fig:DMRG_spin} shows comparable results at zero temperature obtained using DMRG. In both Figs.~\ref{Fig:DQMC_spin} and \ref{Fig:DMRG_spin}, panels {\bf a}-{\bf e} show the spectra obtained from the full $pd$-model while panels~{\bf f}-{\bf j} show the corresponding results obtained from a singleband Hubbard model. In the latter case, we use $t=0.5$~eV $t^{\prime}=0.06t$, $U=2.66$~eV to best fit to the $pd$-model results (see Methods and Supplementary Note 2 for a more detailed discussion). To compare to the singleband model, we present the total spin response here. We stress that both the cluster sizes and simulation temperatures are the same for the two models, and one can directly compare the results shown on both sides of each figure.  

At low-temperature and $\langle \hat{n} \rangle = 1$, the corner-shared cuprates are charge transfer insulators with antiferromagnetic correlations. In this limit, the system's elementary magnetic excitations are spinons, which must be created in pairs. The magnetic excitation spectrum is dominated by  a two-spinon continuum, which has been explicitly observed in, for example, Sr$_2$CuO$_3$ using INS \cite{Walters2009} and RIXS  \cite{SchlappaNature2012, Schlappa2018}. 
The upper and lower boundaries of the two-spinon continuum are given by $\omega_+(q) = \pi J|\sin(qa/2)|$ and $\omega_-(q) = \tfrac{\pi J}{2}|\sin(qa)|$, respectively, where $J$ is the  Cu-Cu antiferromagnetic superexchange energy. 
Our DQMC and DMRG results for both the $pd$- [Fig.~\ref{Fig:DQMC_spin}{\bf a} \& Fig.~\ref{Fig:DMRG_spin}{\bf a}] and singleband [Fig.~\ref{Fig:DQMC_spin}{\bf f} \& Fig.~\ref{Fig:DMRG_spin}{\bf f}] models reflect this behaviour. Specifically, we observe a continuum of magnetic excitations confined within $\omega_\pm(q)$ but with a maximum spectral weight $\omega_m(q)$ [indicated by the cyan points] concentrated at energies near the lower boundary $\omega_-(q)$. This distribution suggests that the low-energy magnetic excitations of the $pd$-model deviate only mildly from the Heisenberg limit, where the low-energy physics is described by the 1D $t$-$J$ model~\cite{PhysRevB.94.205145,Nocera2018SR}. 
By assuming that the locations of the maximum spectral weight in Fig.~\ref{Fig:DQMC_spin}{\bf a} and  Fig.~\ref{Fig:DMRG_spin}{\bf a} correspond to $\omega_-(q)$, we can estimate $J=\frac{2}{\pi}\omega_m/\sin(\frac{\pi}{2})\approx 350$ meV for the multi-orbital model. This value is a little smaller than the value $J \approx 4t^2/U = 376$ meV that one would obtain from the single band Hubbard model in the large $U$ limit. (For this estimate, we have neglected the presence of $t^{\prime}=0.06t$, which would induce a negligible frustration to the system.)

Upon doping $\langle \hat{n}\rangle = 1+x$, the location of the zero-energy mode $q_s$ shifts from $\frac{\pi}{a}$ to $(1-|x|)\frac{\pi}{a}$, consistent with the prior demonstration that the deviation of the wave vector from $\pi/a$ is proportional to hole and electron doping~\cite{PhysRevB.94.205145}. We note that using DQMC, at the largest doping, $|x|=0.2$, the spin excitations at $q_s=0.8\frac{\pi}{a}$ appear to acquire a finite gap at $T = 0.0625$ eV but remain gapless at zero temperature. This behaviour may be an artifact of the Maximum Entropy Method and/or the finite temperature effect of the simulation. 

We observe a hardening of the spin excitation spectrum as the total hole concentration increases. For example, Fig.~\ref{Fig:spin_summary}{\bf a} shows the change of the energy of the maximum intensity $I_\mathrm{max}(q)$ of $S(q,\omega)$ as a function of the hole density at $q=0.4\pi/a$. Here, the solid triangles and circles represent DQMC results, while the open triangles and circles represent DMRG results. Since the DQMC data can be broad due to thermal broadening and the use of the Maximum Entropy method, we also provide approximate error bars, which are estimated as the energy range over which  $S(q,\omega) \ge 0.99I_\mathrm{max}(q)$. The two dashed lines are guides to the eye. For both the $pd$- and downfolded singleband models, we find that the energy of the peak increases with $\langle\hat{n}\rangle$. This observation implies that the spin excitation energy hardens with hole-doping and softens with electron doping. This asymmetric behaviour is opposite to what is observed in the two-dimensional Hubbard model and experimental observations for the 2D cuprates~\cite{JiaNacom2013, LeBlancPRB2019}. Since we observe consistent behaviour in our 1D single- and multi-orbital models, we attribute this difference to the dimensionality of the respective systems. We also note that the spin excitations obtained with DQMC at finite temperature are slightly higher in energy compared to the DMRG results, implying that the spin excitations shift to higher energies as the temperature increases.

We can estimate the effective size of the system's total local moment $m$ from a sum rule that relates it to an integral over the spin structure
\begin{eqnarray}\label{Eq:SumRule}
m=\lim_{\omega_c\rightarrow \infty}\frac{3}{N}\sum_{q}\int_0^{\omega_c} S(q,\omega) d\omega, 
\end{eqnarray}
where $N$ is the number of the unit cells, and the factor of three comes from the sum over the three spin components, which contribute equally because of the unbroken SU(2) symmetry. Since INS experiments typically only access the low energy region, we cut off the integration to $\omega_c=2$ eV here. 
In the limit of strong interactions, the magnetic moment $m=S(S+1)= 0.75$ in the singleband Hubbard model; however, this value will be reduced for finite $U$ due to double occupancy and additional covalency effects in the $pd$-model \cite{Walters2009}. To determine by how much, we evaluated Eq.~\eqref{Eq:SumRule} by integrating our numerical data over binding energies in the interval between 0 and $\omega_c=2$~eV, as shown in Fig.~\ref{Fig:spin_summary}b.
We note that the DQMC results are a little larger than the DMRG results, which is attributed to a broadening effect of the Maximum Entropy method and finite temperature. 
The magnetic moment is about 0.536 at half-filling and zero temperature for the Hubbard model, and it decreases somewhat symmetrically as the system is doped with electrons or holes. These results imply that the effective singleband model has some nonzero reduction due to double occupancy, consistent with previous INS  studies~\cite{LorenzanaPRB,Walters2009,PhysRevLett.93.087202}. For the $pd$-model, we observe that the magnetic moment is about 0.17 smaller compared to the singleband model, both at zero and finite temperature. This result implies that the hybridization with oxygen further reduces the total magnetic moment in cuprate chains. 

Next, we analyze the contribution of each orbital to the total magnetic moment. Figures~\ref{Fig:spin_summary}{\bf c} and \ref{Fig:spin_summary}{\bf d} plot the weight of spin excitations between two Cu orbitals, two O orbitals, and Cu and O orbitals at half filling calculated using  the DQMC and DMRG data. Here, the intraorbital spin excitations between neighbouring Cu and O orbitals are labeled as ``Cu" and ``O", respectively, while interorbital spin excitations between Cu and O are labeled as ``Cu-O". Interestingly, we observe that the spin excitations on the Cu sites have the maximum weight (about $55\%$), and interorbital spin excitations between Cu and O account for $38\%$ of the total excitations, which is nonnegligible.

It is remarkable that the magnetic excitations of the full multi-orbital $pd$-model are well reproduced by our effective singleband model, apart from the overall magnetic moment per unit cell predicted by the two models. This difference is also reflected in the overall intensity of the spin structure factors. 

\subsection*{The Charge Excitations} 

We now examine the charge excitations of the $pd$-model and compare them to the excitations predicted by the singleband model.  Fig.~\ref{Fig:DQMC_charge} plots DQMC results for the dynamic charge structure factor $N(q,\omega)$ obtained from both models, again at $T = 0.0625$~eV and for different hole densities. 
As was the case with our spin results,  $N(q,\omega)$ represents the total charge response and panels {\bf a}-{\bf e} show the excitation spectrum of the $pd$-model while panels~{\bf f}-{\bf j} show the excitation spectrum of the singleband Hubbard model. Fig.~\ref{Fig:DMRG_charge} plots DMRG results for the same models at zero temperature, following the same format.

The charge excitation spectrum of the $pd$-model can be divided into low- and high-energy sectors, with dividing line occurring at $\omega \approx 5$~eV, as shown in Fig.~\ref{Fig:DQMC_charge} and Fig.~\ref{Fig:DMRG_charge}. The high-energy region corresponds to particle-hole excitations from the low-energy bands crossing $E_\mathrm{F}$ to the high-energy oxygen-derived bands, and the UHB observed in the spectral function. The charge excitations in the low-energy region appear as a sharp cosine-like excitation that is gapped at half-filling and gapless in the doped systems. These low-energy features originate from scattering within holon and ZRS bands near the Fermi level. 

At zero temperature, shown in Fig.~\ref{Fig:DMRG_charge}, additional fine structure in the high- and low-energy regions of the spectral function develops. Here, we observe that the low-energy region consists of two distinct branches, one gapped and the other gapless. The gapless excitations are intraband scattering within the holon branch of the doped system and are notably absent in the spectra at half-filling. The gapped excitations are then scattering from the holon band to the remnant of the ZRS band (located at $\omega \in [-3, -1]$ eV in Fig. \ref{fig:Akw_DMRG}{\bf a} and $\omega \in [1, 4]$ eV in Fig. \ref{fig:Akw_DMRG}{\bf c}), respectively. 

When comparing our $pd$-model results to the singleband model, we focus on the low-energy section of the charge excitation spectrum because the high-energy excitations are understandably absent in the singleband model. Overall, we find that the low-energy charge excitations of the $pd$-model are qualitatively well described by the singleband model. The DMRG results show that both the singleband model and the $pd$-model have gapped and gapless charge excitations when $\hat{n}\neq 1$. The gapped excitations of the singleband model originates from the scattering between the holon band and the upper Hubbard band, while this gapped excitations of the $pd$-model come from the scattering between the holon band and the remnant ZRS band. In the DQMC results, due to the broadening of the finite temperature and the Maximum Entropy method, the sharp gapped spectrum is replaced by a broad spectrum, connecting to the spectrum of the gapless excitation.

To better visualize the charge excitations of the DQMC results at low- and high-energy regions, we plot $N(q,\omega)$ of the $pd$-model in Fig.~\ref{Fig:charge_summary} for $\langle \hat{n} \rangle=0.9$, $\langle \hat{n} \rangle=1.1$, $\langle \hat{n} \rangle=0.8$, and $\langle \hat{n} \rangle=1.2$ at $q=\pi/a$. We also include the DMRG results for reference.
We decomposed the DQMC results into three Gaussian functions to distinguish the low- and high-energy charge excitations. The center of these three Gaussian functions coincide with the peak position of the DMRG results.  The summation of these three Gaussian functions matches the original DQMC results very well. Besides, We observe an asymmetry in the intensity of the lowest energy peak between hole and electron doped regimes consistent with the asymmetric orbital content between Cu and O in the undoped ground state.

Even though the low-energy total charge excitations can be described by the singleband model, the orbital-resolved results show a very interesting behaviour, which could help account for the particle-hole asymmetry observed in RIXS experiments~\cite{LeeNPhys2014,Jiaqinpj2020}. 
Figure.~\ref{Fig:charge_orbital} shows the orbital-resolved $N^{\gamma,\gamma^\prime}(q,\omega)$, which is evaluated from both DMRG and DQMC calculations.  Red, cyan, and blue colors in Fig.~\ref{Fig:charge_orbital} represent the charge excitations between Cu-Cu, O-O, and Cu-O, respectively. We observe that the gapless low-energy charge excitation for $\langle \hat{n} \rangle=0.9$ is dominated by the Cu-Cu and Cu-O components, while the similar charge excitation for $\langle \hat{n} \rangle=1.1$ mainly consists of the O-O and Cu-O components. 
In two-dimensional superconducting cuprates, Cu $L_3$-edge RIXS experiments reported a strong charge signal near $q=0$ on the electron doped side, which has been difficult to observe on the hole doped side~\cite{LeeNPhys2014,Jiaqinpj2020}. The missing signal with hole doping may be attributed to the weaker contribution to the charge excitations originating from the Cu-Cu excitations. Our results show that the missing signal can be observed from the O components, consistent with the O $K$-edge RIXS results.

\section*{Discussion}

We have studied the dynamic spin and charge structure factors of a four-orbital $pd$-model relevant for one dimensional cuprate spin chains using numerically exact DQMC, DMRG, and ED. We also compared the two-particle response functions of the $pd$-model against those predicted by an effective downfolded singleband Hubbard model.

Our results show that the singleband Hubbard model can describe the low-energy total spin and charge excitations of the $pd$-model upon hole- or electron-doping;
In the case of the magnetic excitations, we found that the collective excitations harden (soften) with hole (electron) doping. This asymmetry is captured by the singleband Hubbard model with a small positive hopping between next-nearest neighbors. 
We find even richer physics in our orbital-resolved results in the four band $pd$-model. For example, we find that the low-energy spin excitations mainly consist of intraorbital Cu-Cu and interorbital Cu-O components, and we distinguish the dynamical spin behaviors on each orbital, including the effects of electron-electron interactions beyond first principle approaches~\cite{Walters2009}.
Our observations for the collective charge excitations, which have not been widely studied in the context of cuprate physics, are richer. Our results show a gapped charge excitation spectrum in the undoped regime, while gapless excitations develop upon doping with weaker intensity on the hole-doped side than on the electron-doped side at low energy. In the low-energy section, the charge excitations in the electron-doped regime are dominated by the Cu-Cu and Cu-O components while the hole-doped regime by the Cu-O and O-O components. This behaviour reflects the charge transfer insulating nature of the cuprates, where doped holes preferentially reside on the oxygen sublattice while doped electrons reside on the copper sublattice.

Our work has important implications for numerical studies of competing and intertwined orders in one- and two-dimensional cuprates. For example, in the 2D superconducting cuprates, a clear particle-hole asymmetry has been identified in the excitations probed by RIXS experiments~\cite{LeeNPhys2014,Jiaqinpj2020}. Here, a branch of collective modes has been observed in the electron-doped $\mathrm{Nd}_{2-x}\mathrm{Ce}_x\mathrm{CuO}_4$, which has been difficult to observe in hole-doped cuprates. This particle-hole asymmetry of the collective modes is consistent with the behaviours of the charge excitations on the Cu site observed here, which suggests that the explicit inclusion of the oxygen degrees of freedom may be required to capture this physics. It would, therefore, be interesting to extend this study to higher dimensions by considering multi-leg ladders on route towards full 2D models. 

\section*{Methods}
\small
\subsection*{Models}
The Hamiltonian for the four-orbital $pd$-model describing corner-shared cuprates [see Fig.~\ref{Fig:fig1} (a)], written in hole language, is given by
\begin{eqnarray}
H&=&(\epsilon_d-\mu)\sum_{i,\sigma}\hat{n}_{i,\sigma}^d+\sum_{j,\gamma,\sigma}(\epsilon_{p,\gamma}-\mu)\hat{n}_{j,\gamma,\sigma}^{p} \nonumber\\
&+&\sum_{\substack{\langle i,j \rangle \\
 \nonumber
 \gamma,\sigma}} t_{pd}^{ij}\left( d_{i,\sigma}^{\dagger}p_{j,\gamma,\sigma}^{\phantom\dagger} + h.c.  \right) 
+\sum_{\substack{\langle j,j^\prime \rangle\\ \gamma,\gamma^\prime,\sigma}} t_{pp}^{jj^\prime} p^{\dagger}_{j,\gamma,\sigma} p^{\phantom\dagger}_{j^\prime,\gamma^\prime,\sigma}
\nonumber \\
&+&U_d\sum_{i}\hat{n}_{i,\uparrow}^{d} \hat{n}_{i,\downarrow}^{d} 
+U_p\sum_{j,\gamma} \hat{n}_{j,\gamma,\uparrow}^{p}\hat{n}_{j,\gamma,\downarrow}^{p} \nonumber \\
&+&U_{pd}\sum_{\substack{\langle i,j, \gamma\rangle \\ \sigma,\sigma^\prime}} \hat{n}_{i,\sigma}^{d} \hat{n}_{j,\gamma,\sigma^\prime}^p.\label{Eq:Hpd}
\end{eqnarray}
Here, $\langle \cdots \rangle$ denotes a sum over nearest neighbor orbitals; $d^{\dagger}_{i,\sigma}$  and $p^\dagger_{j,\gamma,\sigma}$ creates a hole with spin $\sigma$ ($=\uparrow,~\downarrow$) on the $i^{\mathrm{th}}$ Cu $3d_{x^2-y^2}$ orbital and the $j^\mathrm{th}$ O $2p_\gamma$ ($\gamma=x,\pm y$) orbital, respectively; $\epsilon_d$ and  $\epsilon_{p,\gamma}$ are the onsite energies; $\hat{n}_{i,\sigma}^{d}$ and  $\hat{n}_{j,\gamma,\sigma}^p$ are the number operators for the Cu $3d_{x^2-y^2}$ orbital and O $2p_\gamma$ orbital, respectively; $t_{pd}^{ij}$ and $t_{pp}^{jj^\prime}$ are the nearest-neighbor Cu-O and O-O hopping integrals, whose phase factors are drawn in Fig.~\ref{Fig:fig1}(a); $U_d$ and $U_p$ are the onsite Hubbard interactions on the Cu and O orbitals, respectively, and $U_{pd}$ is the nearest-neighbor Cu-O Coulomb repulsion; Finally, 
$\mu$ is the chemical potential, which is adjusted to control the hole density in our DQMC simulations. 

Throughout, we adopt parameters determined from LDA calculations and comparisons to experiments~\cite{NeudertPRB2000, WohlfeldPRB}. Specifically, we set (in units of eV) $\epsilon_d=0$, $\epsilon_{p,x}=3$, $\epsilon_{p,y}=3.5$, $|t_{(p,x)d}|=1.5$, $|t_{(p,y)d}|=1.8$, $|t_{pp}|=0.75$, $U_d=8$, $U_p=4$, and $U_{pd}=1$. 

We map the low-energy spin properties of the four-orbital $pd$-model to a singleband $t-t^\prime$ Hubbard model. To remain consistent with the $pd$-model, our singleband Hubbard model is also written in the hole language, and is given by
\begin{eqnarray}
H=-\mu \sum_{i,\sigma} \hat{n}_{i,\sigma} + 
\sum_{i,\sigma} t_{i,j}c_{i,\sigma}^{\dagger}c_{ij,\sigma}^{\phantom\dagger}+U\sum_{i}\hat{n}_{i,\uparrow}\hat{n}_{i,\downarrow}.
\end{eqnarray}
Here, $t_{i,j} = t$ and $t^\prime$ are the nearest- and next-nearest-neighbor hopping integrals (we set all longer range hopping to zero). We adopt $t=0.5$ eV based on the analysis in  Ref.~\citenum{Nocera2018SR}. 
To determine the remaining parameters, we adjusted $t^\prime$ and $U$ to fit the dynamic magnetic susceptibility of the $pd$-model and found that $U=2.66$ eV and $t^\prime=0.06t$ produce the best agreement between these two models (see Supplementary Note 2). 

\subsection*{Determinant Quantum Monte Carlo}
The details of the DQMC algorithm applied to the multi-orbital Hubbard models can be found in Ref.~\citenum{Shaozhithesis}. DQMC works in the canonical ensemble, where the expectation value of an observable $\hat{O}$ is given by $\langle \hat{O} \rangle = Z^{-1} \mathrm{Tr}[\hat{O}\mathrm{e}^{-\beta H}]$, where $Z=\mathrm{Tr}[\mathrm{e}^{-\beta H}]$ is the partition function and $\beta$ is the inverse temperature.

To study the model's excited state properties, we measured the imaginary-time dynamic magnetic $\chi_s$ and charge $\chi_c$ susceptibilities. They are given by
\begin{eqnarray}\label{Eq:ChiS}
\chi_s^{\gamma,\gamma^\prime}(q,\tau)=\langle \hat{S}^{\gamma,z}_q(\tau) \hat{S}^{\gamma^\prime,z}_{-q}(0) \rangle
\end{eqnarray}
and 
\begin{eqnarray}\label{Eq:ChiC}
\chi_c^{\gamma,\gamma^\prime}(q,\tau)=\langle \hat{n}^{\gamma}_q(\tau) \hat{n}^{\gamma^\prime}_{-q}(0) \rangle, 
\end{eqnarray}
where $\hat{S}^{\gamma,z}_q(\tau)=\hat{n}^{\gamma}_{q,\uparrow}(\tau)-\hat{n}^{\gamma}_{q,\downarrow}(\tau)$ and $\hat{n}^{\gamma}_q(\tau)=\hat{n}^{\gamma}_{q,\uparrow}(\tau)+\hat{n}^{\gamma}_{q,\downarrow}(\tau)$. Here, $\gamma$ ($\gamma^\prime$) is the orbital index, and $\hat{n}^{\gamma}_{q,\sigma}$ and $\hat{S}_q^{\gamma,z}$ are the Fourier transforms of the local density and spin-$z$ operators.

To compare to the singleband model, we calculate the total spin and charge responses, which are given by
\begin{eqnarray}
\chi_s({q,\tau})=\langle \hat{S}^{z}_q(\tau) \hat{S}^{z}_{-q}(0)\rangle
\end{eqnarray}
and
\begin{eqnarray}
\chi_c({q,\tau})=\langle \hat{n}^{z}_q(\tau) \hat{n}^{z}_{-q}(0)\rangle,
\end{eqnarray}
where $\hat{S}^z_q=\sum_{i,\gamma}\mathrm{e}^{iq r_{i,\gamma}}\hat{S}^{z}_{i,\gamma}$ and $\hat{n}^z_q=\sum_{i,\gamma}\mathrm{e}^{iq r_{i,\gamma}}\hat{n}_{i,\gamma}$.
Here, $r_{i,\gamma}$ represents the position of the orbital $\gamma$. Besides, we also calculate the spin and charge responses between Cu (O) and O sites, where the operator on the O site is given by $\hat{O}=\hat{O}_{p_x}+\hat{O}_{p_y}+\hat{O}_{p-y}$.

To examine the spectral properties, we then used the method of the maximum of entropy~\cite{FuchsPRE2010} to analytically continue the imaginary-time susceptibilities to the real frequency axis. The same analytic continuation method has been used by some of the authors to study the 2D cuprates, and reasonable results were obtained~\cite{LeBlancPRB2019}. The dynamical spin and charge structure factors are calculated by the fluctuation-dissipation theorem, which simplifies to
\begin{eqnarray}
S(q,\omega)=\frac{\mathrm{Im}\chi_s(q,\omega)}{1-\mathrm{e}^{-\beta\omega}}
\end{eqnarray}
and
\begin{eqnarray}
N(q,\omega)=\frac{\mathrm{Im}\chi_c(q,\omega)}{1-\mathrm{e}^{-\beta\omega}}.
\end{eqnarray}

The primary drawback to DQMC is the Fermion sign problem~\cite{LohPRB}, which limits the range of accessible temperatures and Hubbard interactions. In general, we have found that the sign problem is alleviated in 1D systems \cite{ShaozhiPRB2018}, and the smallest value of the sign we obtained in our simulations is about 0.78, much larger than the sign value of the 2D three-orbital $pd$-model~\cite{Huang1161}.

All of our DQMC calculations are performed on  $N=20$ chains (for a total of 40 orbitals total in the multi-orbital case). The simulation temperature was held at $T=0.0625$ eV for both the single- and multi-orbital calculations.

\subsection*{Density Matrix Renormalization Group}

The DMRG\cite{White92,White93} calculations were carried out with the correction-vector method\cite{PhysRevB.60.335} using the Krylov decomposition\cite{PhysRevE.94.053308}, 
as implemented in the DMRG++ code\cite{alvarez0209}. 
This approach requires real-space representations for the dynamical structure factors in Eqs.~\eqref{Eq:ChiS} and \eqref{Eq:ChiC}, which can be found in  Ref. \citenum{Nocera2018b}. Here, we calculated the response functions for $N=20$ unit cell long chains and open boundary conditions, which corresponds to total system sizes of $N$ and $4N+1$ orbitals for the single- and multi-orbital cases, respectively.  We kept up to $m=1000$ DMRG states to maintain a truncation error below $10^{-7}$ and introduced a spectral broadening in the  correction-vector approach fixed at $\eta=0.1$~eV for both the single- and multi-band calculations.

\small
\bibliography{main}

\section*{Acknowledgements}
We thank C. D. Batista and T. A. Maier for useful discussions and comments on the manuscript. This work was supported by the U.S. Department of Energy, Office of Basic Energy Sciences, Materials Sciences and Engineering Division. 
S.J. acknowledges support from the Scientific Discovery through Advanced Computing (SciDAC) program funded by the U.S. Department of Energy, Office of Science, Advanced Scientific Computing Research and Basic Energy Sciences, Division of Materials Sciences and Engineering. A. N. acknowledges support from the Max Planck-UBC-UTokyo Center for Quantum Materials and Canada First Research  Excellence Fund (CFREF) Quantum Materials and Future Technologies Program of the Stewart Blusson Quantum Matter Institute (SBQMI), and the Natural Sciences and Engineering Research Council of Canada (NSERC). U. K. acknowledges support from the US DOE NNSA under Contract No.
89233218CNA000001 through the LDRD Program. This research used resources of the Compute and Data Environment for Science (CADES) at the Oak Ridge National Laboratory, which is supported by the Office of Science of the U.S. Department of Energy under Contract No. DE-AC05-00OR22725. This work also used computational resources and services  provided by Compute Canada and Advanced Research Computing at the University of British Columbia. 

\section*{Author contributions} 
S.L. performed DQMC calculations. A.N. performed DMRG calculations. S.L. and U.K. performed exact diagonalization calculations. S.L. and S.J. developed the DQMC code. S.J. supervised the project. All authors contributed to analyzing the data and writing the manuscript.

\section*{Additional information}

\textbf{Competing interests}: The authors declare no competing interests.\\

\noindent \textbf{Supplementary Information} accompanies this paper at \url{Insert_link}.\\

\noindent \textbf{Code Availability}: DQMC and ED codes can be downloaded at \url{https://github.com/sli43/one_dimensional_four_orbital_pd_model}. The DMRG code can be downloaded at \url{https://github.com/g1257/dmrgpp/}. \\

\noindent \textbf{Data Availability}: The data that support the findings of this study are available from the corresponding
authors on reasonable request.

\newpage

\newpage

\newpage

\newpage

\newpage
\begin{figure*}[ht]
\centering
\includegraphics[width=\textwidth]{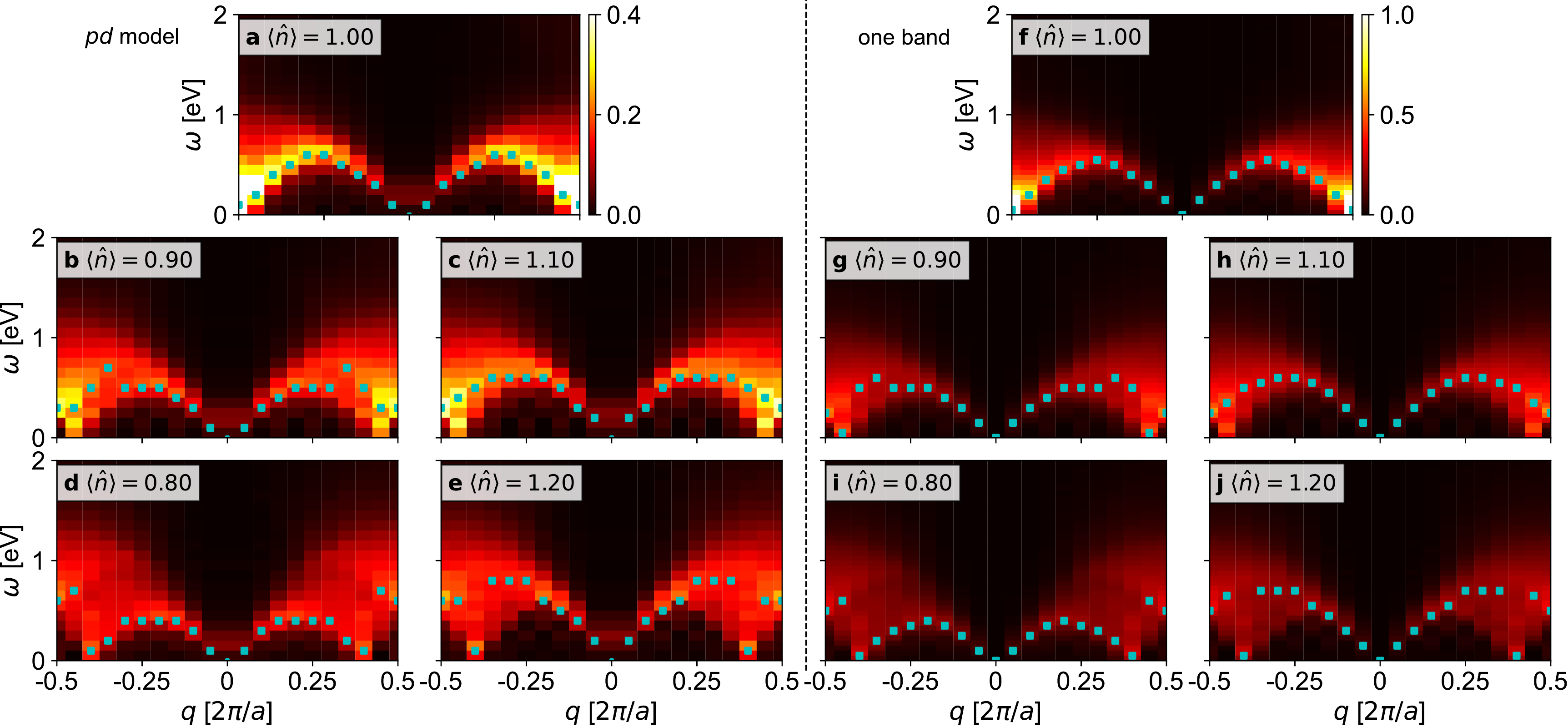}
\caption{\textbf{Zero temperature DMRG results for the total dynamical spin structure factor of the multi-orbital $\mathbf{pd}$- and singleband Hubbard models.} Panels {\bf a}-{\bf e} show the dynamical spin structure $S(q,\omega)$, obtained from DMRG simulations of the full multi-orbital $pd$-model at various fillings, as indicated. Panels {\bf f}-{\bf j} show corresponding results for DMRG simulations of the singleband Hubbard model. Both sets of results were obtained using chains with $N = 20$ unit cells.}\label{Fig:DMRG_spin}
\end{figure*}

\newpage
\begin{figure*}[ht]
\centering
\includegraphics[width=0.6\columnwidth]{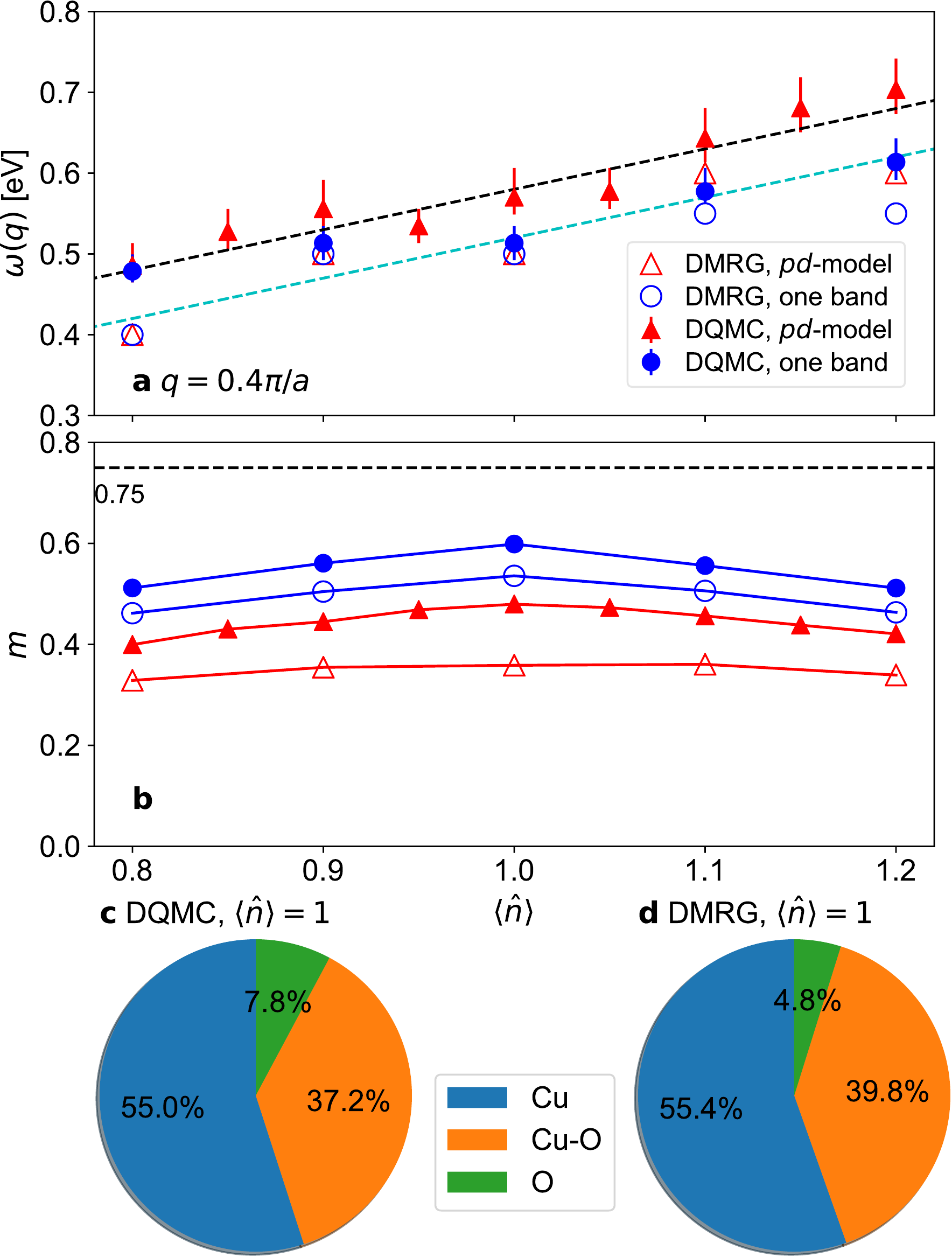}
\caption{\textbf{The evolution of the magnetic excitations with doping.}  
{\bf a} The shift in energy of the maximum of $S(q=0.4\pi/a,\omega)$ as a function of doping in the multi-orbital $pd$- and singleband models. The dashed lines are guides to the eye. {\bf b} The magnetic moment $m$ obtained from integrating  $S(q,\omega)$ from $\omega \in [0,2]$ eV. Results are shown for DQMC 
at $T = 0.0625$ eV and DMRG at zero temperature. Panels {\bf c} and {\bf d} show the weight of the spin excitations between neighbour Cu, O, and Cu and O orbitals. Spin excitations between neighbor Cu and O orbitals are labeled as Cu and O, respectively. Interorbital spin excitations between Cu and O are labeled as Cu-O.}\label{Fig:spin_summary}
\end{figure*}

\newpage
\begin{figure*}[ht]
\centering
\includegraphics[width=\textwidth]{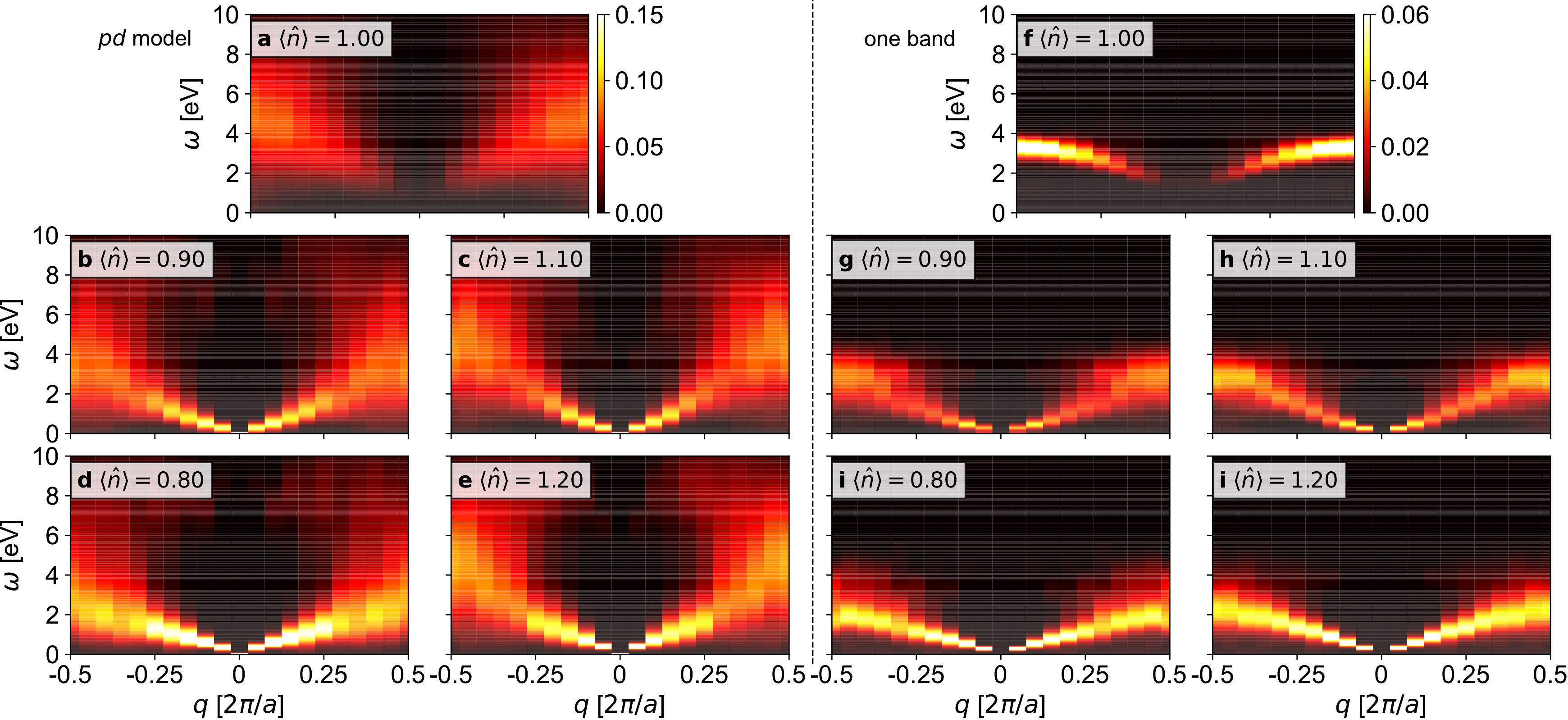}
\caption{\textbf{Finite temperature DQMC results for the total dynamical charge structure factor of the multi-orbital $\mathbf{pd}$-and singleband Hubbard models.} Panels {\bf a}-{\bf e} show the dynamical charge structure $N(q,\omega)$, obtained from DQMC simulations of the full multi-orbital $pd$-model at various fillings, as indicated. Panels {\bf f}-{\bf j} show corresponding results for DQMC simulations of the singleband Hubbard model. Both sets of results were obtained using chains with $N = 20$ unit cells and at temperature $T = 0.0625$ eV.}
\label{Fig:DQMC_charge}
\end{figure*}

\clearpage

\begin{figure*}[ht]
\centering
\includegraphics[width=\textwidth]{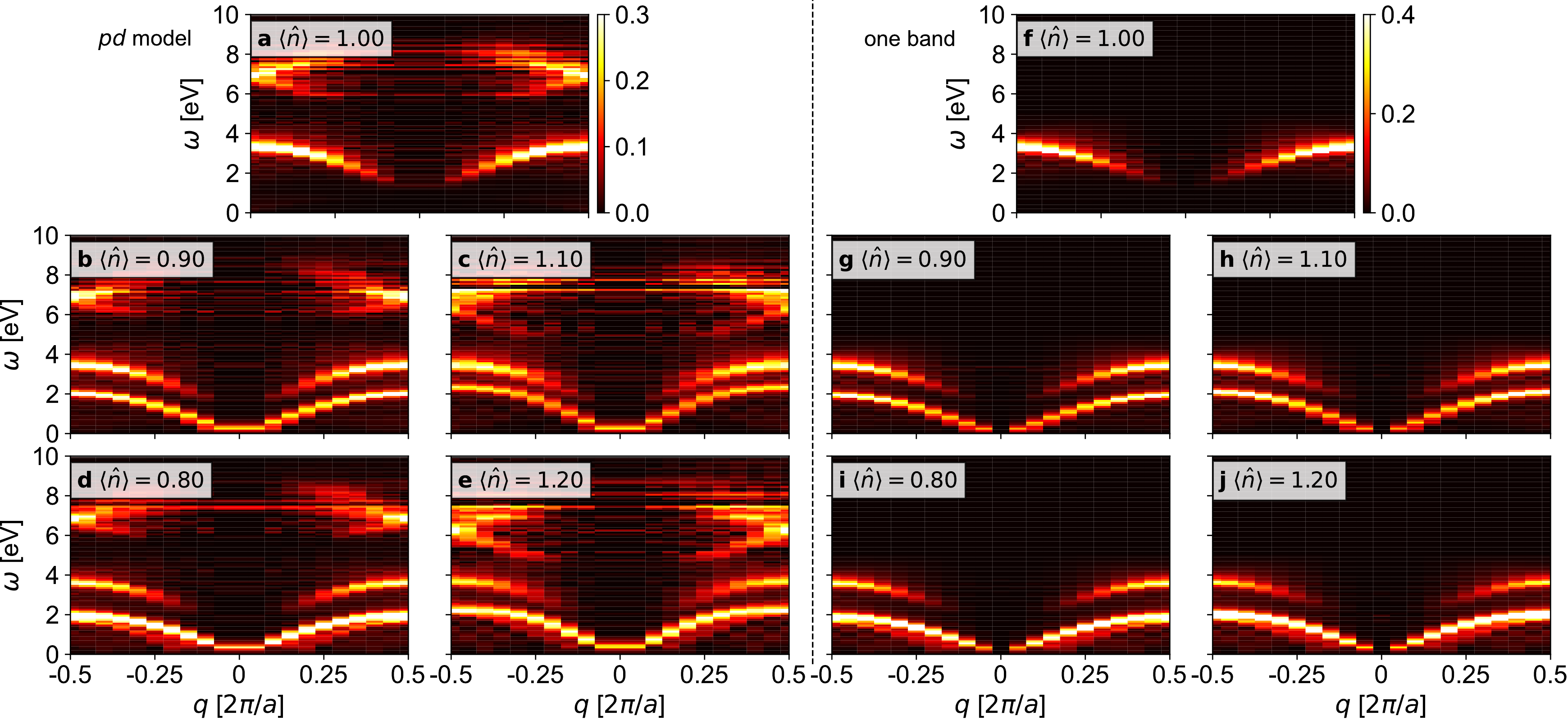}
\caption{\textbf{Zero temperature DMRG results for the total dynamical charge structure factor of the multi-orbital $\mathbf{pd}$- and singleband Hubbard models.} Panels {\bf a}-{\bf e} show the dynamical charge structure factor $N(q,\omega)$, obtained from DMRG simulations of the full multi-orbital $pd$-model at various fillings, as indicated. Panels {\bf f}-{\bf j} show corresponding results for DMRG simulations of the singleband Hubbard model. Both sets of results were obtained using chains with $N = 20$ unit cells.}\label{Fig:DMRG_charge}
\end{figure*}

\newpage
\begin{figure*}[ht]
\centering
\includegraphics[width=0.6\columnwidth]{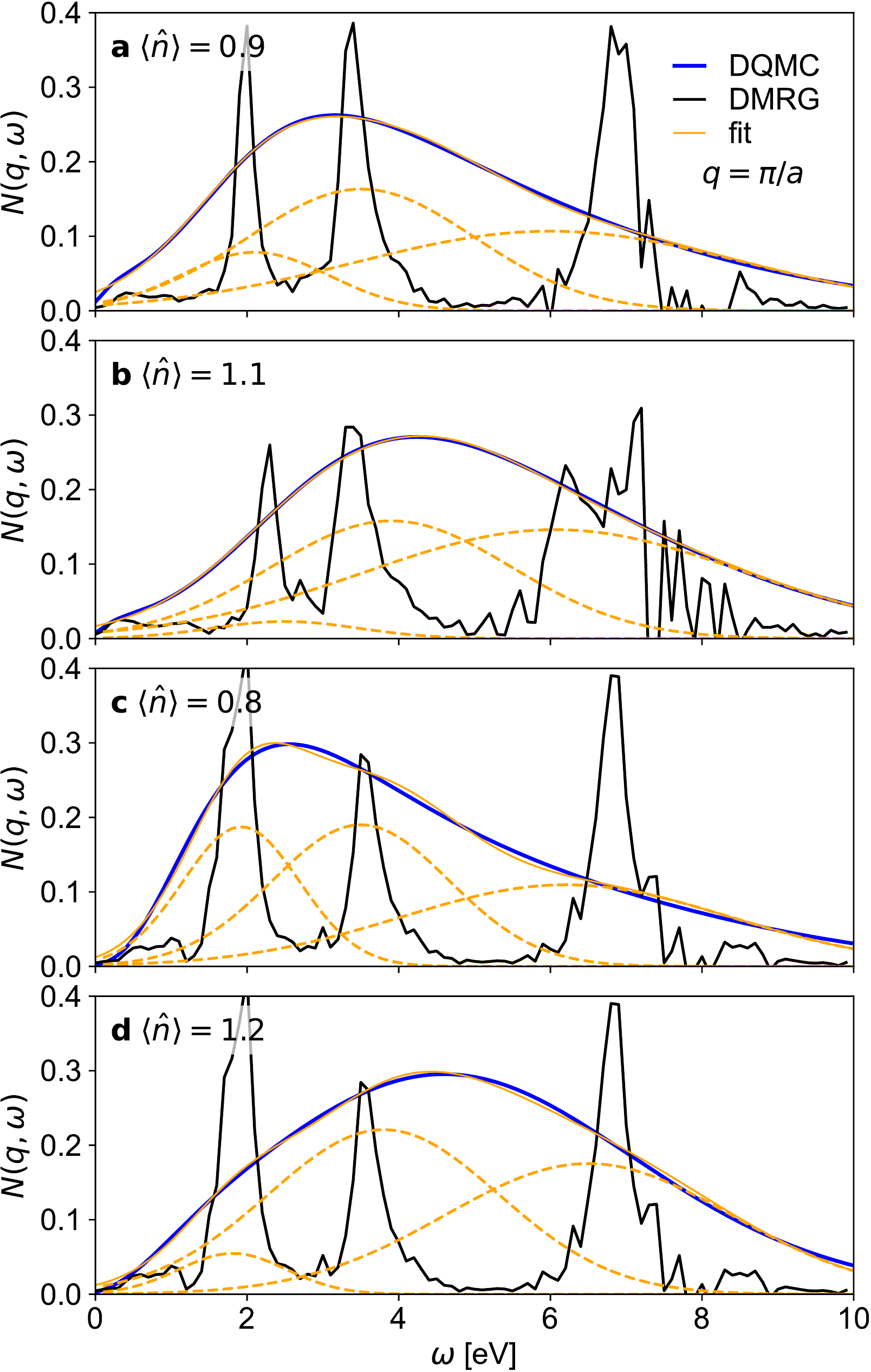}
\caption{\textbf{A comparison of the charge excitations between DQMC and DMRG results.} 
Panels {\bf a}-{\bf d} show the dynamical charge structure factor $N(q,\omega)$ at $q = \pi/a$ for $\hat{n}=0.9$, $\hat{n}=1.1$, $\hat{n}=0.8$, and $\hat{n}=1.2$, respectively. 
The DQMC spectra (blue line) have been fit with a set of Gaussian distributions, whose energies correspond well with the main peaks observed in the DMRG data (black line).}\label{Fig:charge_summary}
\end{figure*} 

\newpage
\begin{figure*}[ht]
\centering
\includegraphics[width=\textwidth]{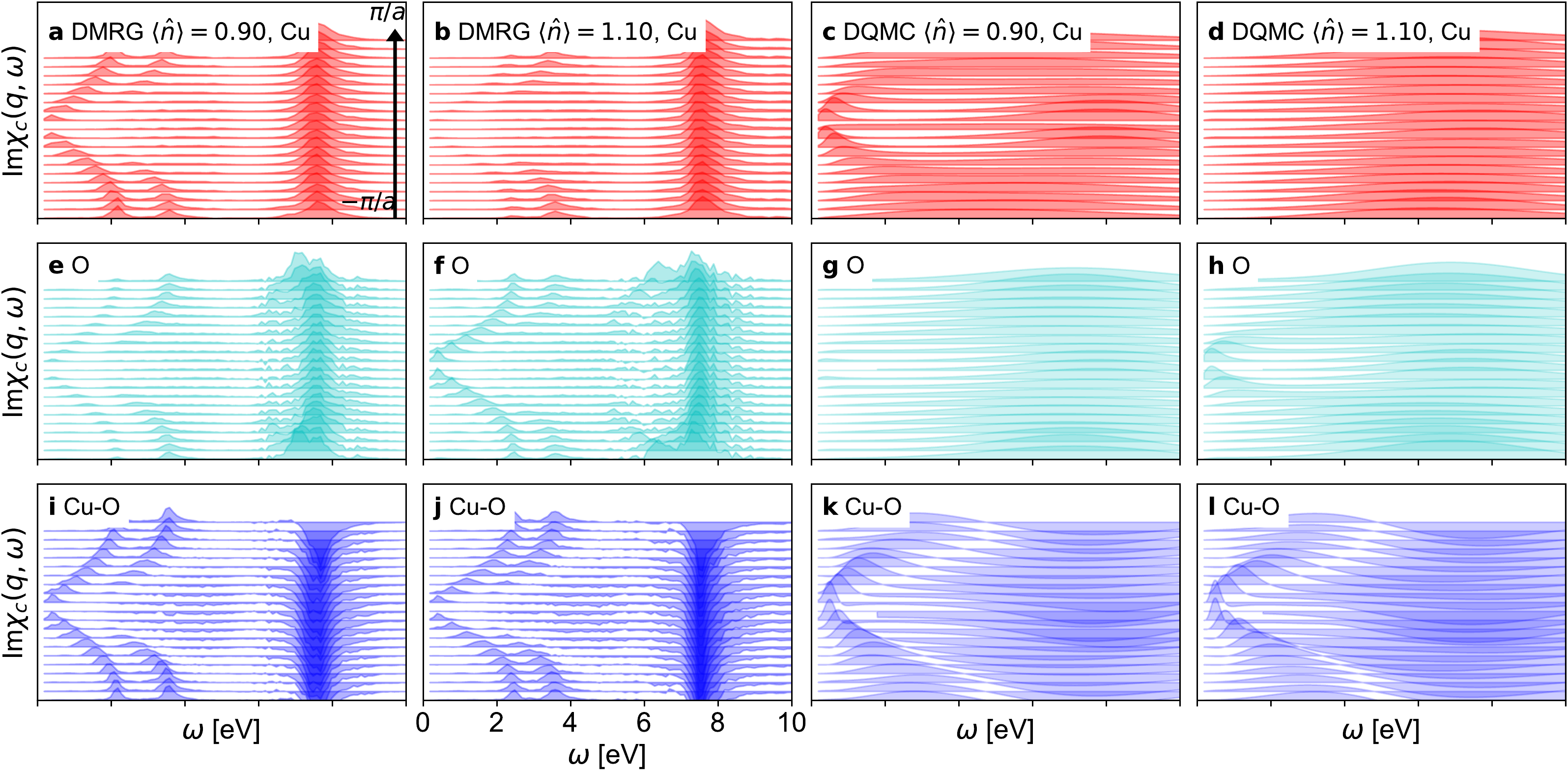}
\caption{\textbf{The orbital-resolved dynamic charge structure factor computed using DMRG and DQMC.} Results are shown for $\langle \hat{n} \rangle=0.9$ and $1.1$ holes/Cu. Red, cyan, and blue colors represent Cu $d_{x^2-y^2}$, O $2p$, and Cu-O components.}
\label{Fig:charge_orbital}
\end{figure*}

\end{document}